\documentclass[12pt]{article}
\usepackage{graphicx}
\renewcommand{\baselinestretch}{1.5}

\renewcommand{\headheight}{20mm}

\begin{document}

\begin{center}

{\Large\bf ABOUT FORMATION OF QUASIBOUND STATES AND ORDERED
STRUCTURES IN DENSE PLASMA }\\[1cm]

T.S. Ramazanov$^{a)}$, M.A. Bekenov, N.F. Baimbetov\\ [10mm]

\it\begin{tabular}{ll}

 & SRIETP Al Farabi Kazakh  National University, \\
     & Tole bi, 96a, 480012, Almaty, Kazakhstan \\[5mm]
\end{tabular}

\vspace*{5mm}

{\bf Abstract:}
\end{center}

The possibility of formation of quasibound states and ordered
structures in dense  plasma is investigated. The effective
potentials of dense plasma are used. On the basis of these models
the condition of the ordered structures formation in the system is
obtained. It is shown that in the dense classical plasma the
quasibound states form  whereas the  ordered structures form in
the dense semiclassical  plasma.

\vspace*{3cm} \noindent $^a)$ E-mail address: ramazan@physics.kz

\newpage
\renewcommand{\headheight}{-20mm}

{\bf I. INTRODUCTION}\\

\vspace*{3mm}

It is well known$^{1-4}$ that a classical Coulomb system form
several ordered structures. As shown in Refs. 1 and 2, the
strongly coupled one-component plasma will crystallize into a
centered cubic form at the values of the coupling parameter
$\Gamma \gg 1$. Here, $\Gamma = e^2/(ak_B T)$; $e$ is the
electrical charge; $a = (3/4\pi n)^{1/3}$ is the average distance
between  particles (Wigner - Seitz radius); $n$ is the number
density of particles; $k_B$ and $T$ are the Boltzmann constant and
the temperature of plasma, respectively.

In previous papers$^{3,4}$, we studied the structural and
thermodynamic properties of the two-component model plasma. The
formation of the ``near ordering'' and anomalous increasing of the
particle correlation radius as a consequence of microstructure
changes in plasma have been demonstrated by Monte Carlo simulation
method. The ``hexatic''- like structures which characterized of
liquid crystals were found in Refs. 5,6. In Refs. 7,8 the
properties of the shell structures of finite one-component plasma
clouds have been investigated. The study of ordered structures is
especially important in connection with the rapid development of
dusty plasma. The ordered structures that form in a strongly
coupled plasma have been first analyzed theoretically in Ref. 9
and have come to be known as plasma crystals. The plasma crystals
of dust particles were found also in a partially ionized plasma in
a radio-frequency discharge$^{10,11}$, in the DC glow
discharge$^{12}$ and in the UV - induced dusty plasmas under
microgravity$^{13}$.

It is now recognized that the plasma crystals are new material for future technologies.
Therefore, the investigation of several ordered structures in  Coulomb systems plays an
important role.

\vspace*{2mm}

{\bf II. INTERACTION  MODELS}\\ \vspace*{3mm}

In this work we consider  a fully ionized, dense (classical and
semiclassical) hydrogen plasma. The number density is considered
in the range $n = n_e = n_i \sim (10^{19} \div 2\times10^{25})$
cm$^{-3}$, and the temperature domain is $T \sim (5\times10^{4}
\div 10^{6})$ K.

According to Ref. 14, we can separate a two types of dense plasma.
For example at $\theta \gg 1$ we have a classical dense plasma.
Here $\theta = k_{B} T / E_{F}$, $k_{B}$ denotes the Boltzmann
constant, $T$, $E_{F}$ are a temperature and the Fermi energy,
respectively. In this case, the quantum mechanical diffraction and
symmetry effects are negligible except in short-range collisions.
When $\theta<0,1$, the electrons are in the state of full Fermi
degeneracy. In the intermediate region between these cases
($\theta\leq 1$), the electrons are partially degenerate and we
have a dense semiclassical plasma.

Due to the well known long range character of the Coulomb
interaction between particles in plasma, the correlation effects
play an important role for dense plasma. Consequently, for the
dense plasma, the simultaneous interaction of a great number of
particles should be taken into account. In Ref. 15, an
integro-differential equation for the effective pair potential has
been derived on the basis of Bogolyubov's chain equation. This
equation takes into account  simultaneous correlations of $N$
particles. In the case of three particle approximations the
expression for the effective potential for  dense plasma
is$^{15,16}$:

\begin{equation} \label{claspot}
\Psi(R) = \frac{\gamma}{R} \, e^{-R} \, \frac{ 1 + \gamma f(R)/2 }
  { 1 + c(\gamma) } .
\end{equation}
Here  $f(R)=(e^{-\sqrt{\gamma} R}-1)(1-e^{-2R})/5$
and $R=r/r_D$, where $r_D$ is the Debye screening length. The potential is
expressed in terms of the thermal energy, $\Psi(R)=\Phi(R)/k_B T$, and
$\gamma=e^2/(r_D k_B T)$ is a nonideality plasma parameter. $c(\gamma)$ is
the correction coefficient for differrent values of $\gamma$:\,
$c(\gamma) = -0.008617 + 0.455861\gamma - 0.108389\gamma^{2} + 0.009377\gamma^{3}$.

For the semiclassical plasma, the effective potential of
Kelbg-Deutsch-Yukhnovskii$^{17,18}$ is usally used:

\begin{eqnarray} \label{KDY}
\Phi_{\alpha\beta}(r) &=& \frac{Z_\alpha Z_\beta e^2}{r} \left[ 1
- \exp( -r / \lambda_{\alpha\beta}) \right] + \\ & &
\delta_{\alpha\beta} \delta_{e \alpha} k_B T \ln(2) \exp\left( -
\frac{r^2}{\pi \ln(2)\lambda_{ee}^2}\right) ,\nonumber
\end{eqnarray}
where $\lambda_{\alpha\beta} = \hbar/(2\pi\mu_{\alpha\beta}k_B
T)^{1/2}$ is the thermal de Broglie wavelength;
$\mu_{\alpha\beta}$ is the reduced mass of electrons and ions.
Notice that the effective potential (\ref{KDY}) does not account
for screening effects in plasma and behaves like the Coulomb
potential for $r\rightarrow\infty$. Therefore, in Ref. 19, the
effective semiclassical potential was obtained by applying the
spline approximation to potential (\ref{KDY}) and the numerical
solution of following Equation$^{19}$:

\begin{equation} \label{psigam}
\Delta \Psi - 3 \Gamma \Psi = \pm 3 \Gamma \Psi^2 ,
\end{equation}
with the boundary conditions
\begin{equation} \label{boundary}
\Psi \mid_{R \rightarrow 0} = \Gamma / R \; ; \quad
\Psi \mid_{R \rightarrow \infty} = 0
\end{equation}
at the intersection point.

The effective potential $\Psi(R)$ is expressed in units of $k_B T$
and $R=r/a$; $\Delta$ is the Laplace operator. In Equation
(\ref{psigam}) the minus and plus signs correspond to the
interaction of particles with equal and opposite charges,
respectively. The spline approximation was performed at the
intersection point of Eq.(\ref{KDY}) and the numerical solution of
(\ref{psigam}) -- (\ref{boundary}). This potential contains
quantum diffraction and symmetry effects at short distances as
well as screening effects for large distances (see Fig.1).

\vspace*{2mm}

{\bf III. ORDERED STRUCTURES IN DENSE PLASMA }\\

\vspace*{3mm}

Following$^{20}$, let us introduce a potential function for the
mathematical description of the system state. This function takes
into account the collective interaction between the components of
statistical system:
\begin{equation} \label{potfun}
U(\vec{r}, \vec{v}, t) = \int \Phi(\vec{r}, \vec{v}, \vec{r}\,', \vec{v}\,' t)f(\vec{r}\,',
\vec{v}\,' t)d\vec{r}\,'d\vec{v}\,'.
\end{equation}
Here $\Phi(\vec{r}, \vec{v}, \vec{r}\,', \vec{v}\,' t)$ is the
potential  describing the interaction of particles in the system;
$\vec{a}$, $\vec{v}$ are a coordinate and a velocity of particles,
respectively. \\

If many-particles effects are considered (e.g. three particle effects), the expression
for $U(\vec{r}, \vec{v}, t)$ becomes non-linear:

\begin{eqnarray} \label{potfun3}
U(\vec{r}, t) = \int \Phi_{12}(\left|\vec{r}-\vec{r}\,'\right|)f(\vec{r}\,',
\vec{v}, t)d\vec{r}\,'d\vec{v} + \int \int \Phi_{123}(\left|\vec{r}-\vec{r}\,'\right|,
\left|\vec{r}-\vec{r}\,''\right|,\left|\vec{r}\,'-\vec{r}\,''\right|)\\
f(\vec{r}\,', \vec{v}\,', t)f(\vec{r}\,'', \vec{v}\,'', t)d\vec{r}\,'d\vec{v}\,'\vec{r}\,
''d\vec{v}\,''. \nonumber
\end{eqnarray}
In the case of pair central interaction between particles we can
written as
\begin{equation} \label{pairpot}
 \Phi(\vec{r}, \vec{v}, \vec{r}\,', \vec{v}\,' t) = \Phi(\left|\vec{r}-\vec{r}\,'\right|).
\end{equation}
Taking into consideration Eq.(\ref{pairpot}), the expression for
potential function is obtained:
\begin{equation} \label{potfun2}
U(\vec{r},t) = \int
\Phi(\left|\vec{r}-\vec{r}\,'\right|)f(\vec{r}\,', \vec{v},
t)d\vec{r}\,'d\vec{v}.
\end{equation}

Considering the effective potentials and the field of collective
interactions, we can obtain the following set of equations for the
potential function $U$ and plasma particles density function
$\rho$$^{20}$:
\begin{eqnarray} \label{Uro}
U(\vec{r}) = C \int \Phi_{eff}(\left|\vec{r}-\vec{r}\,'\right|)e^{-U(\vec{r}\,')/
\theta}d\vec{r}\,'\\
\rho(\vec{r}) = C \exp\left[ - \frac{1}{\theta} U(\vec{r})\right], \nonumber
\end{eqnarray}
where $\theta = 1/k_B T$; $C$ is the constant.

We consider$^{20}$ the occurrence problem of space-periodical
solution for Eq.(\ref{Uro}). In this case, the expression for
$U(\vec{r})$ can be written in the following form:
\begin{equation} \label{phi}
\varphi(\vec{r}) = \lambda \int
\Phi_{eff}(\left|\vec{r}-\vec{r}\,'\right|)e^{\varphi(\vec{r}\,')}d\vec{r}\,'
,
\end{equation}
where $\varphi=-U/\theta$; $\lambda=-C/\theta$. Let us that at
$\lambda=\lambda_0$ and $\varphi=\varphi_0$ we have a violation of
space-uniform solution. Let $\lambda=\lambda_0 + \varepsilon$,
$\varphi=\varphi_0 + u(\vec{r})$ and the equation for $u(\vec{r})$
is written as:
\begin{equation} \label{phiou}
\varphi_0 + u(\vec{r}) = (\lambda_0 + \varepsilon) \int
\Phi_{eff}(\left|\vec{r}-\vec{r}\,'\right|)e^{\varphi_0 +
u(\vec{r}\,')}d\vec{r}\,' ,
\end{equation}
After expanding the element of integration as a power series in
$u(\vec{r})$, we have the following expression:
\begin{equation} \label{lu}
L[u(\vec{r})] = u(\vec{r})- \lambda_0^{*} \int
\Phi_{eff}(\left|\vec{r}-\vec{r}\,'\right|)u(\vec{r}\,')d\vec{r}\,'
,
\end{equation}
where $\lambda_0^{*} = \lambda_0e^{\varphi_0}$; $\sigma(0) =
4\pi\int_0^{\infty}\Phi(s)s^2ds$.\\ We consider the following
linear equation
\begin{equation} \label{u}
u(\vec{r}) = \lambda_0e^{\varphi_0} \int
\Phi_{eff}(\left|\vec{r}-\vec{r}\,'\right|)u(\vec{r}\,')d\vec{r}\,'
.
\end{equation}
According to the occurrence problem of periodical structures from
homogeneous medium, the solution for $u(\vec{r})$ is sought as
$u(\vec{r}) = Ae^{i\vec{k}\vec{r}}$. After inserting this solution
into the Eq. (\ref{u}) and integration, we have the following
expression
\begin{equation} \label{exp}
e^{i\vec{k}\vec{r}} = \lambda_0e^{\varphi_0}\cdot
e^{i\vec{k}\vec{r}}\cdot 4\pi
\int_0^{\infty}\Phi_{eff}(s)\frac{\sin(ks)}{ks}s^2ds.
\end{equation}

Consequently, in order that ordered structures in the system can
form, it is necessary that the periodical solution of the
following equation of nonlocal statistical mechanics to
exist$^{20}$:
\begin{equation} \label{condition}
\frac{\theta}{\rho} = -4\pi \int_0^\infty \Phi_{eff}(s)\cdot
\frac{\sin(ks)}{ks}s^2 ds \equiv \sigma(k).
\end{equation}
In the present work, the behavior of the function $\sigma(k)$ is
investigated for the above mentioned models of dense plasma; here
$k$ is a wave number.

The model of  dense classical plasma is characterized by the fact
that for the low limit of integral (\ref{condition}) we must take
a certain non-zero value of distance $R_0 = r/r_D$. In this case,
$R_0$ is minimal distance between particles. In Figure 2, the
results of numerical analysis of expression (\ref{condition}) on
the basis of effective potential (\ref{claspot}) are shown. Notice
that, with increasing of $\gamma$, on the dependencies of
$\sigma(k)$ some non-monotonic behavior are observed. In this
case, the increase of $\gamma$ is caused by the density increase
(the decreasing of $r_D$). It is well known that for a dense
plasma the number of particles in Debye sphere $N_D<1$. It is
obtained also that functions $\sigma(k)$ form local minimums when
 $R_0 \sim 1$ ($r\approx r_D$). These facts can be interpreted as the formation
 of quasibound states in dense classical   plasma. The term
 "quasibound" states$^{21,22}$ denotes that the energy
 $\varepsilon$ of electron-proton pairs is
 $-e^2/a<\varepsilon\leq0$. Because the size of "quasibound"
 states is greater than $a$, it is considered that the electron
 moves over a part of elliptical orbit in the field of the nearest
 proton, thereupon the electron passes into the field of another
 proton, and so on.

For  the model of dense semiclassical  plasma, the problem of
introduction of minimal distance is absent because the
pseudopotential (full line in Fig.1) is limited for $r=0$ and it
is strongly screened when $r\rightarrow \infty$. The results of
numerical analysis of Eq. (\ref{condition}) are represented in
Figures 3,4 and 5 for different values of the density parameter
$r_s = a/a_B$ and the coupling parameter $\Gamma =
Z_{\alpha}Z_{\beta} e^2/(ak_B T)$. Here, $a$ is the average
distance between particles and $a_B$ is the Bohr radius. It is
shown that no oscillations on dependencies of $\sigma_{ee}(k)$ for
electron-electron component are observed. It is probably connected
with the fact that electrons do not form complexes (structures)
between themselves because of their little masses. The functions
$\sigma_{ep}$ and $\sigma_{pp}$  have pronounced oscillations that
can be considered as formation of ordered structures.\\ In Refs.
2,23 it has been demonstrated that in the intermediate region
($1\leq\Gamma\leq150$) the plasma can be considered as a ordered
structures of ions. According to this model, ions form a
cubic-like space structures due to the strong interactions between
particles. Since we have a chaotic motion in the system, these
structures are realized "on the average". Therefore, the energy of
this structure is greater then Madelung's energy of simple
centered cubic lattice. These configurations are called as
"ordered structures".

Consequently in dense semiclassical hydrogen plasma, the
``picture'' of particles distribution can be as follows. Protons
are distributed with a certain order and they are surrounded by
electron clouds. The  ``proton-electron'' system can be considered
as a quasi-particle.

This work was supported by Ministry of Sciences of Kazakhstan, Grant MN-2.1/98 and by
INTAS-RFBR, Grant No.95-1335. Our thanks are due to Yu.L.Klimontovitch, V.E.Fortov and
A.P.Nefedov for many stimulating discussions.


\newpage

\large
\renewcommand{\baselinestretch}{1.0}

\newpage
\vspace*{3cm}
\bigskip
\centerline{\bf  Figure captions }
\bigskip
\vspace{1cm}

\noindent Figure 1:
 Effective electron-proton potential for dense semiclassical hydrogen plasma at $r_s=1$
 and $\Gamma=2$ (solid line). Triangles denote the numerical solution of
 Eq.\ (\ref{psigam}) for a dense, classical plasma$^{15}$  which
 accounts for higher-order screening effects. The dashed line denotes
 potential$^{17,18}$ which shows quantum corrections at short distances.
 The dot-dashed line: DH potential.

\vspace{2mm}
\noindent Figure 2:
 Reduced $\sigma^* = \sigma/(r_D^3k_BT)$ functions for dense classical plasma
 at $R_0 = 2.$

\vspace{2mm}
\noindent Figure 3:
 Reduced $\sigma_{ee}^* = \sigma_{ee}/(a^3k_BT)$ functions between electrons of
 dense semiclassical  hydrogen plasma at $r_s = 5.$

\vspace{2mm}
\noindent Figure 4:
 Reduced $\sigma_{ep}^* = \sigma_{ep}/(a^3k_BT)$ electron-proton functions  of
 dense semiclassical  hydrogen plasma at $r_s = 2.$

\vspace{2mm} \noindent Figure 5: Reduced $\sigma_{pp}^* =
\sigma_{pp}/(a^3k_BT)$ functions between protons of dense
semiclassical  hydrogen plasma at $r_s = 1.$

\newpage

\vspace{3cm}

\begin{figure}[h]
\includegraphics[width=13cm,height=12cm]{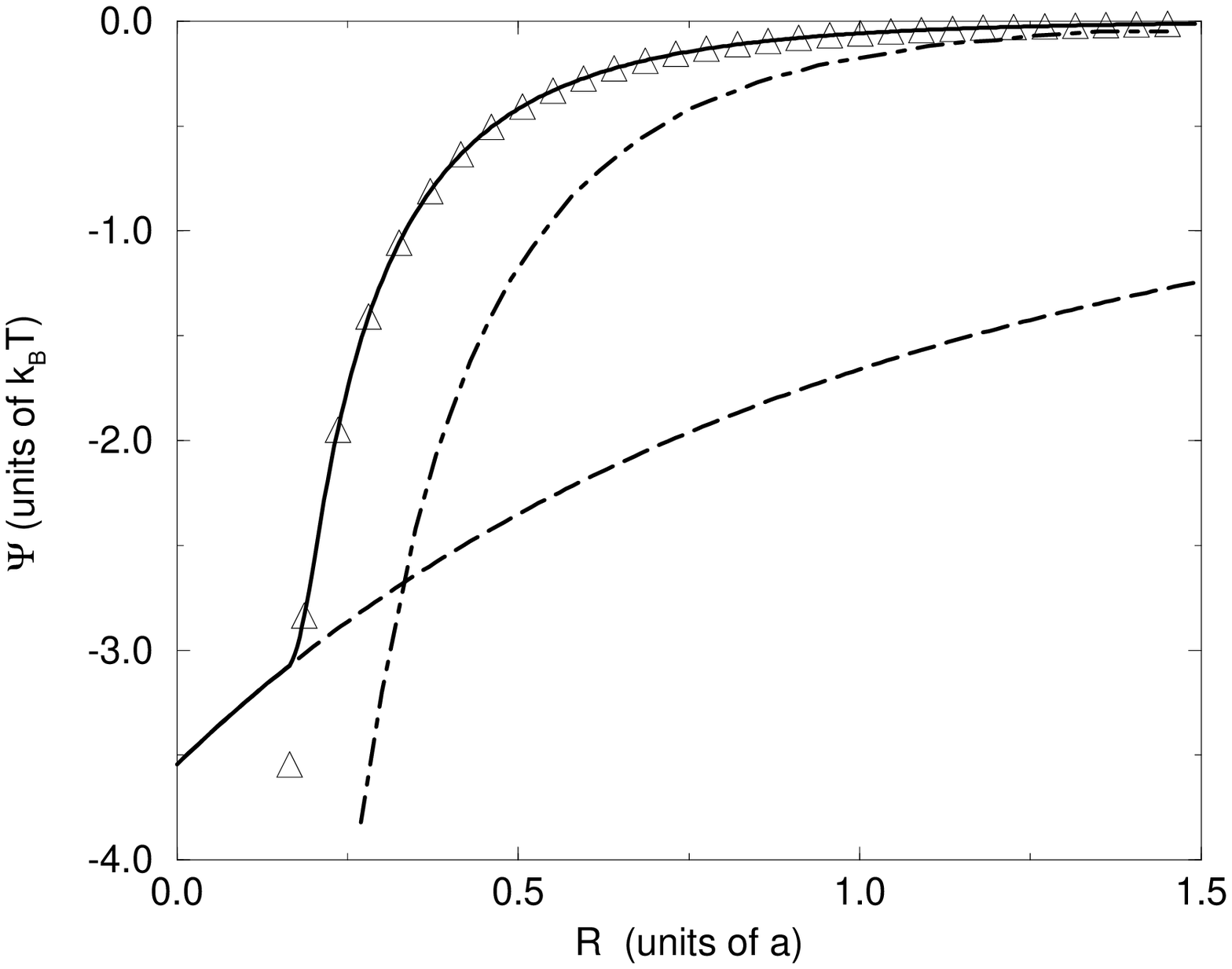}
\end{figure}
\vspace{2cm}
T.S.Ramazanov, M.A.Bekenov, N.F.Baimbetov\\
\noindent Figure 1:
{\small Effective electron-proton potential for dense
 semiclassical hydrogen plasma at $r_s=1$
 and $\Gamma=2$ (solid line). Triangles denote the numerical solution of
 Eq.\ (\ref{psigam}) for a dense, classical plasma$^{15}$  which
 accounts for higher-order screening effects. The dashed line denotes
 potential$^{17,18}$ which shows quantum corrections at short distances.
 The dot-dashed line: DH potential.}

\newpage

\vspace*{3cm}

\begin{figure}[h]
\includegraphics[width=11cm,height=11cm]{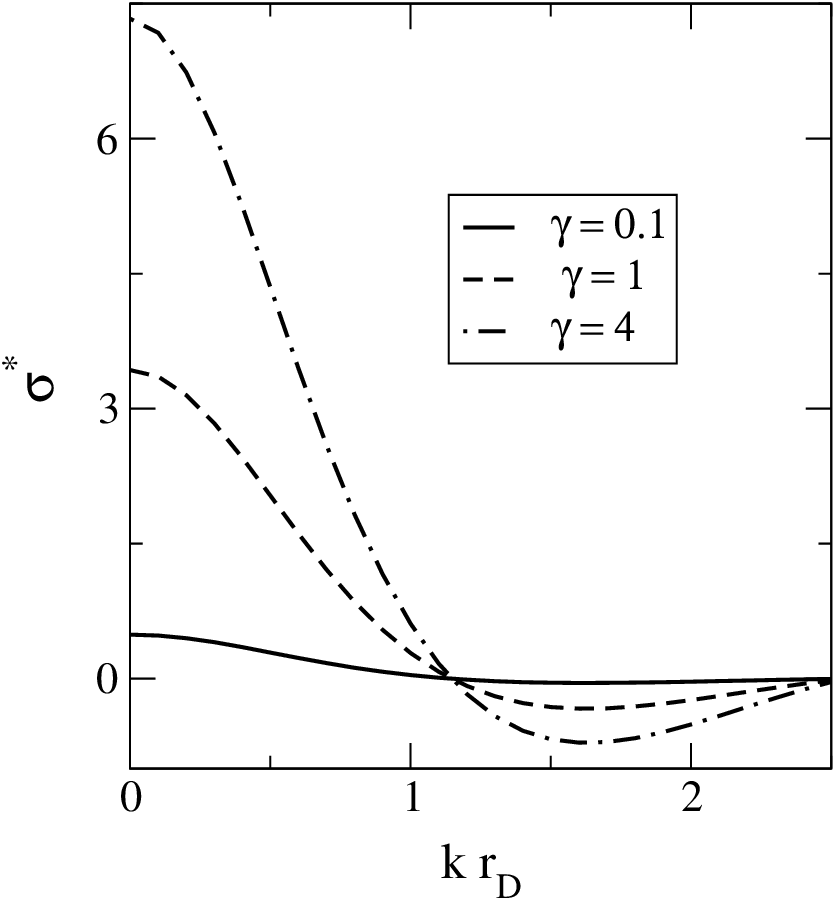}
\end{figure}
\noindent T.S.Ramazanov, M.A.Bekenov, N.F.Baimbetov\\
\noindent Figure 2: {\small Reduced $\sigma^* =
\sigma/(r_D^3k_BT)$ functions for dense classical  plasma at $R_0
= 2.$  }

\newpage

\vspace*{3cm}

\begin{figure}[h]
\includegraphics[width=11cm,height=11cm]{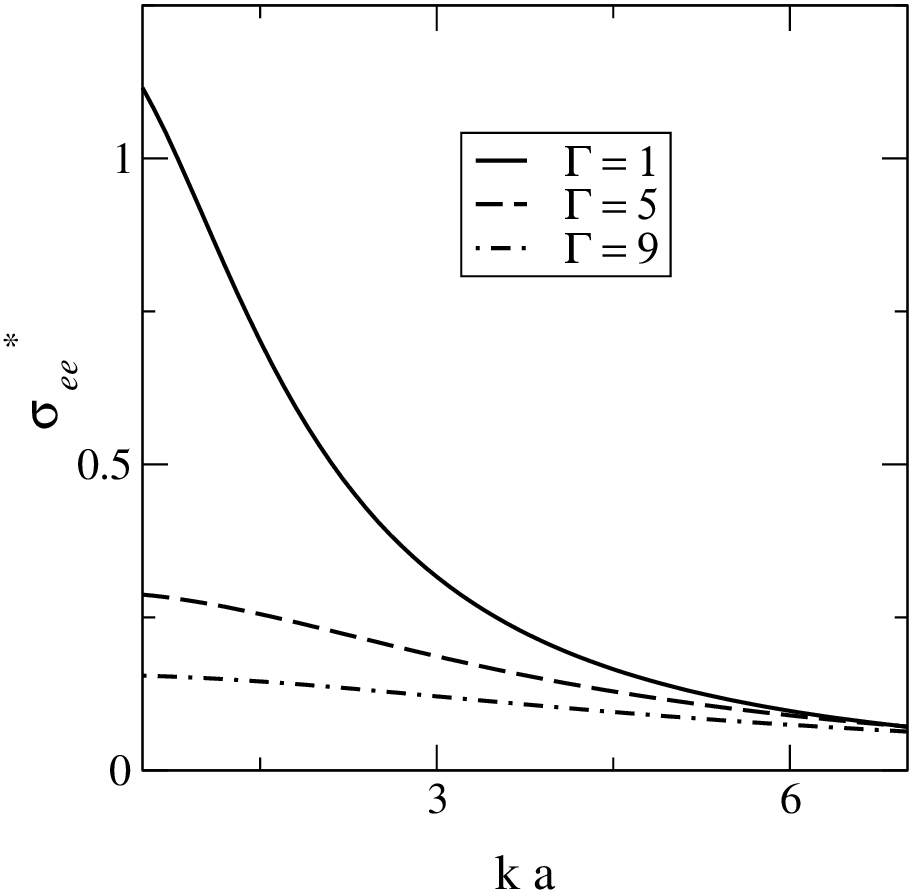}
\end{figure}
\noindent T.S.Ramazanov, M.A.Bekenov, N.F.Baimbetov\\
\noindent Figure 3: {\small Reduced $\sigma_{ee}^* =
\sigma_{ee}/(a^3k_BT)$ functions between electrons of dense
semiclassical  hydrogen plasma at $r_s = 5.$}

\newpage

\vspace*{3cm}

\begin{figure}[h]
\includegraphics[width=11cm,height=11cm]{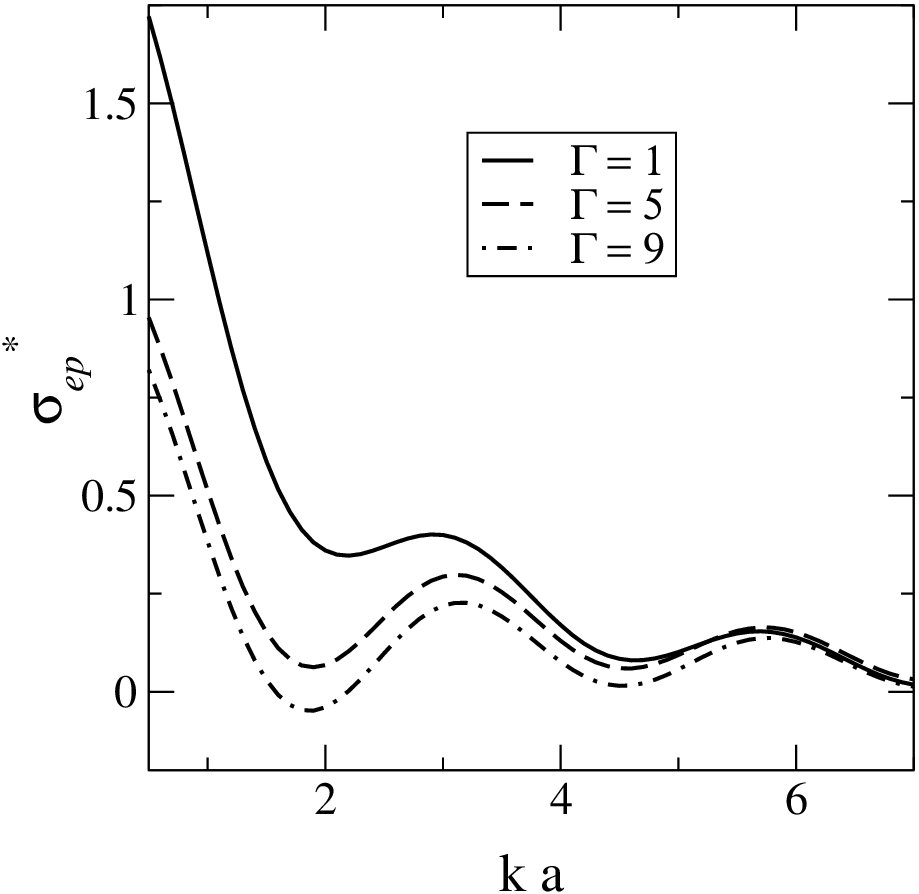}
\end{figure}
\noindent T.S.Ramazanov, M.A.Bekenov, N.F.Baimbetov\\
\noindent Figure 4: {\small Reduced $\sigma_{ep}^* =
\sigma_{ep}/(a^3k_BT)$ electron-proton functions of dense
semiclassical  hydrogen plasma at $r_s = 2.$}

\newpage

\vspace*{3cm}

\begin{figure}[h]
\includegraphics[width=11cm,height=11cm]{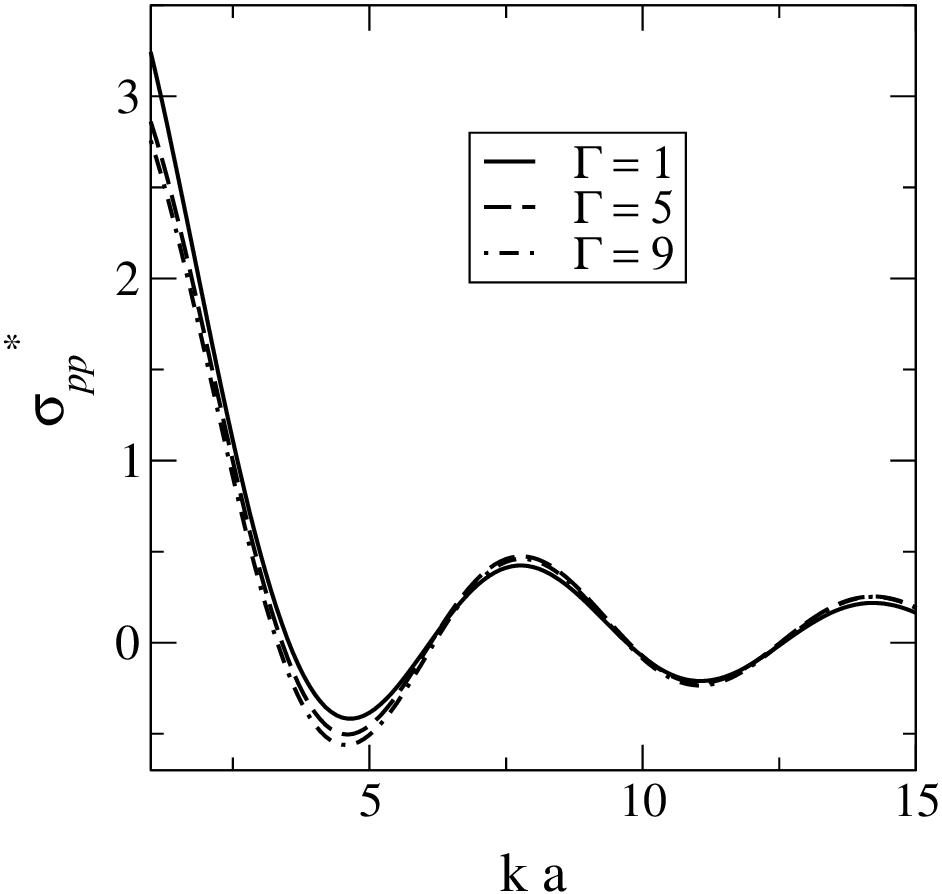}
\end{figure}
\noindent T.S.Ramazanov, M.A.Bekenov, N.F.Baimbetov\\
\noindent Figure 5: {\small Reduced $\sigma_{pp}^* =
\sigma_{pp}/(a^3k_BT)$ functions between protons of dense
semiclassical  hydrogen plasma at $r_s = 1.$ }


\normalsize


\begin{thebibliography}{99}


\bibitem{Brush} S.G.Brush, H.L.Sahlin and E.Teller, J.Chem.Phys., {\bf 45}, 2102 (1966).

\bibitem{Pollock} E.L.Pollock and J.P.Hansen, Phys.Rev., A {\bf 8}, 3110 (1973).

\bibitem{brs90} F.B. Baimbetov, T.S. Ramazanov, N.B.Shaltykov, Teplofizika vysokikh
temperatur, {\bf 28}, 595 (1990).

\bibitem{ts95} T.S. Ramazanov, High Temperature, {\bf 33}, 153 (1995).

\bibitem{Rahman} A.Rahman and J.P.Schiffer, Phys.Rev.Letters, {\bf 57}, 1133 (1986).

\bibitem{Gilbert} S.L.Gilbert, J.J.Bollinger and D.J. Wineland, Phys.Rev.Letters,
{\bf 60}, 2022 (1988).

\bibitem{Schiffer} J.P.Schiffer, Phys.Rev.Letters, {\bf 70}, 818 (1993).

\bibitem{Dubin} D.H.Dubin and J.P.Schiffer, Phys.Rev., E {\bf 53}, 5249 (1996).

\bibitem{Ikezi} H.Ikezi, Phys. Fluids, {\bf 29}, 1764 (1986).

\bibitem{Chu} J.H.Chu and I.Lin, Phys.Rev.Letters, {\bf 72}, 4009 (1994).

\bibitem{Thomas} H.Thomas, G.E.Morfill, V.Demmel, Phys.Rev.Letters, {\bf 73}, 652 (1994).

\bibitem{VE97} V.E.Fortov, A.P.Nefedov, V.M.Torchinsky, {\it et al.}, Physics Letters, A
{\bf 229}, 317 (1997).

\bibitem{VE98} V.E.Fortov, A.P.Nefedov, O.S.Vaulina, {\it et al.}, J. Exp. Theor. Phys.,
{\bf 87}, 1087 (1998).

\bibitem{ichi82} S.Ichimaru, Rev. Mod. Phys., {\bf 54}, 1017 (1982).

\bibitem{bnr95} F.B. Baimbetov, Kh.T. Nurekenov and T.S. Ramazanov,
 Phys. Lett. A {\bf 202}, 211 (1995).

\bibitem{bnr96} F.B. Baimbetov, Kh.T. Nurekenov and T.S. Ramazanov,
 Physica A {\bf 226}, 181 (1996).

\bibitem{KELBG}
G. Kelbg, Ann. Phys. (Leipzig) {\bf 12}, 219 (1963); {\it ibid.} {\bf 14},
 354 (1964).

\bibitem{dgm81} C. Deutsch, M.M. Gombert and H. Minoo, Phys. Rev. A {\bf 23},
 924 (1981).

\bibitem{bbr95} F.B. Baimbetov, M.A. Bekenov and T.S. Ramazanov, Phys. Lett. A
 {\bf 197}, 157 (1995).

\bibitem{Vlasov} A.A.Vlasov, Non-local statistical mechanics (Nauka, Moscow,
1978) 320.

\bibitem{homkin} V.S.Vorobiov, A.L.Homkin, Teplofizika vysokikh
temperatur, {\bf 15}, 188 (1977).

\bibitem{fortov} V.E.Fortov, I.T.Iakubov, Physics of Nonideal
Plasma (Hemisphere Publishing, New York, 1990) 556.

\bibitem{hansen} J.P.Hansen, Phys.Rev., A {\bf 8}, 3096 (1973).







\end{thebibliography}
\end{document}